# Temporal and spatial chaos in the Kerr-AdS black hole in an extended phase space

Bing Tang[*]

*Department of Physics, Jishou University, Jishou 416000, China*

ABSTRACT

Based on the Melnikov method, we investigate chaotic behaviors in the extended thermodynamic phase space for a slowly rotating Kerr-AdS black hole under temporal and spatial perturbations. Our results show that the temporal perturbation coming from a thermal quench of the spinodal region in the phase diagram may cause the temporal chaos only when the perturbation amplitude is above a critical value, which involves the angular momentum $J$. Under the spatial perturbation, however, it is found that the spatial chaos always occurs, which is independent of the perturbation amplitude.

## 1. Introduction

Since Hawking [1] firstly proved that black holes can radiate thermal energy, more and more attention has been focused on the study of black hole thermodynamics [2-12]. Several black holes have been indicated to reveal some important properties in different dimensions, which can give new understanding of the fundamental problem associated with their critical behaviors. In particular, both charged [3] and rotating asymptotically AdS back holes [4,9] allow a first order phase transition between small-black-hole and large-black-hole, which is analogous to the transition between liquid and gas in the Van der Waals fluid. Theoretically, this analogy can be extended to more general cases by analyzing thermodynamics in an extended phase space of the black holes, where the cosmological constant is regarded as the thermodynamic pressure and its conjugate quantity is identified with thermodynamic volume. Based on these considerations, Kubizňák *et al.* have investigated the $P-V$ criticality of charged AdS black holes[13]. They have perfectly clarified the analogy between charged AdS black holes and the liquid–gas system, which has been first observed by

[*] Corresponding author.
E-mail addresses: bingtangphy@jsu.edu.cn

Chamblin *et al.*[3]. Subsequently, $P-V$ criticality in the extended phase space of black holes has become an important direction of the study on the phase transition [14-17].

Chaos is a universal physical phenomenon in those nonlinear dynamical systems. Due to the intrinsic nonlinearity of Einstein's general relativity theory, the chaotic behavior is an important dynamics characteristic for such relativistic systems [18-22]. In the past few decades, more and more attention has been focused on the chaotic behavior in the black hole background [23-37]. As a relatively early work in this field, Letelier and Vieira have proved that the particles moving along timelike geodesics of the Schwarzschild black hole can present a chaotic motion when a particular class of gravitational perturbations are introduced [23]. Furthermore, some works have showed that the motion of a particle is also chaotic when it approaches very near to the black hole horizon [26-33]. Recently, Chabab *et al.* [34] have studied chaotic phenomenon in the context of black hole thermodynamics and phase transitions by means of the Melnikov method [38]. The deep connection between the charged AdS black hole and the Van der Waals fluid has been established for the first time. Their results can help to understand the correlation between the chaotic behavior and the $P-V$ diagram of black holes. Subsequently, this study has been extended to other black holes, including charged or neutral Gauss-Bonnet AdS black holes [35], the Born-Infeld-AdS black hole [36], and the charged dilaton-AdS black hole [37]. Unfortunately, little attention has been paid to chaos phenomena in the Kerr-AdS black hole within the extended phase space.

Gunasekaran *et al.* have proved that a slowly rotating Kerr-AdS black hole is analogous to the Van der Walls liquid-gas system[17]. When the temperature of the black hole is below the critical value, a region of the violated stable equilibrium can appear. By the use of Maxwell's equal area law, Zhao *et al.* have found an isobar for the $P-V$ diagram of the Kerr-AdS black hole, which corresponds to the real two phase coexistence line [39]. Based on this fact, we predict that chaos phenomena may exist in the Kerr-AdS black hole within the extended phase space. To prove this prediction, we apply the Melnikov method to investigate temporal and spatial chaos phenomena of the slowly rotating Kerr-AdS black hole in extended thermodynamic phase space. Firstly, the main thermodynamic characteristics of the Kerr-AdS black hole is reviewed. Secondly, a small temporal perturbation in the spinodal region of the Kerr-AdS black hole thermodynamic phase space is introduced. Then, we derive the Melnikov function for the homoclinic orbit and display when the temporal chaos can occur in the spinodal region of the thermodynamic phase space. Lastly, we recompute the Melnikov function for homoclinic or heteroclinic orbit and probe spatial chaotic behavior. More details will be displayed in the following sections.

## 2. Thermodynamics of the Kerr-AdS black hole in an extended phase space

In this section, let us give an overview on thermodynamics of a four-dimensional AdS rotating black hole. It can be described by the following Kerr-AdS metric [9,40]

$$ds^2 = -\frac{\Delta}{\rho^2}\left(dt - \frac{a\sin^2\theta}{\Xi}d\varphi\right)^2 + \frac{\rho^2}{\Delta}dr^2 + \frac{\rho^2}{\Sigma}d\theta^2$$
$$+ \frac{\Sigma\sin^2\theta}{\rho^2}\left(adt - \frac{r^2+a^2}{\Xi}d\varphi\right)^2$$

(1)

with

$$\rho^2 = r^2 + a^2\cos^2\theta, \quad \Xi = 1 - \frac{a^2}{l^2}, \quad \Sigma = 1 - \frac{a^2}{l^2}\cos^2\theta, \quad \Delta = (r^2+a^2)(1+\frac{l^2}{r^2}) - 2mr. \quad (2)$$

where $l$ is the AdS radius. The associated thermodynamic quantities are as follows[9,17,41]:

$$S = \frac{\pi(r_+^2 + a^2)}{\Xi}, \quad T = \frac{r_+(1 + \frac{a^2}{l^2} + 3\frac{r_+^2}{l^2} - \frac{a^2}{r_+^2})}{4\pi(r_+^2 + a^2)}, \quad \Omega_H = \frac{a\Xi}{r_+^2 + a^2}. \quad (3)$$

The mass $M$ and the angular momentum $J$ are respectively connected with parameters $m$ and $a$:

$$M = \frac{m}{\Xi^2}, \quad J = \frac{am}{\Xi^2}. \quad (4)$$

In the extended phase space, the cosmological constant $\Lambda = -\frac{3}{l^2}$ can be interpreted as a thermodynamic pressure $P$, which has the following relation

$$P = -\frac{1}{8\pi}\Lambda = \frac{3}{8\pi}\frac{1}{l^2}. \quad (5)$$

In this case, the first law of black hole thermodynamics and Smarr formula are respectively written as

$$\delta M = T\delta S + \Omega_H \delta J + V\delta P, \quad (6)$$

$$\frac{M}{2} = TS + \Omega_H J - VP. \quad (7)$$

Here, $V$ is the thermodynamic volume conjugate to $P$. In the present work, we focus on a slowly rotating Kerr-AdS black hole. Thus, after neglecting all higher order terms of $J$, the equation of state can be written as [4,39]

$$P = \frac{T}{v} - \frac{1}{2\pi v^2} + \frac{48J^2}{\pi v^6}, \quad (8)$$

where $v$ is the specific volume satisfying the relation

$$v = 2\left(\frac{3V}{4\pi}\right)^{1/3} = 2r_+ + \frac{12}{r_+(3r_+^2 + 8\pi r_+^4 P)}J^2. \quad (9)$$

Eq. (8) suggests that some $P-v$ critical behaviours really exist. The "real" phase diagram of the slowly rotating Kerr-AdS black hole has been investigated in the extended phase space via Maxwell's equal area law [39]. There exists a second order transition, i.e., between large and small black hole phase transition, which is analogous to between the gas and the fluid phase transition in the Van der Waals system. This phase transition occurs at the following critical point

$$P_c = \frac{1}{36\pi\sqrt{10}}\frac{1}{J}, v_c = 2\times 90^{\frac{1}{4}}\sqrt{J}, T_c = \frac{90^{\frac{3}{4}}}{225\pi}\frac{1}{\sqrt{J}}. \quad (10)$$

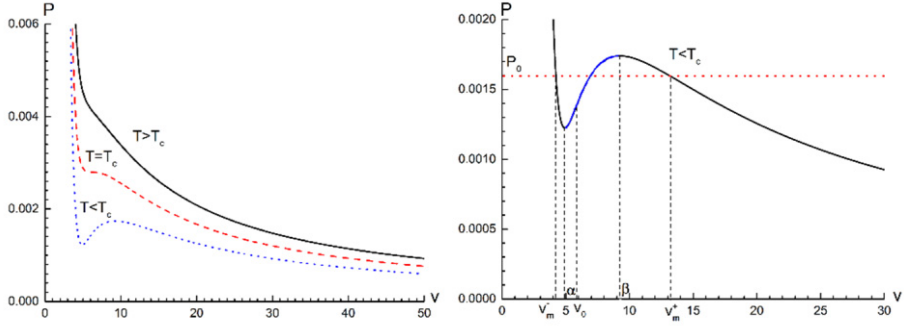

Fig. 1. The $P-v$ diagram of the Kerr-AdS black hole for different temperatures. The angular momentum parameter is chosen as $J=1$. On the right panel with the case of $T<T_c$, it is obvious that the curve is divided into three regions, which contains two stable regions (black lines) and one region (blue line). The red dot line represents the coexisting line of the small black hole (with specific volume $v_m^-$) and the large black hole (with specific volume $v_m^+$) with transition pressure $P_0$, which satisfies Maxwell's equal area law.

In Fig.1, we show the $P-v$ diagram corresponding to the Kerr-AdS black hole. We can clearly see that a small-large black hole phase transition exists in the case of $T<T_c$. Especially, an example of this transition is directly presented on the right panel of Fig. 1. Notice that the $P-v$ curve contains two stable regions and one unstable region. Two stable regions are $v\in[0,\alpha]$ and $v\in[\beta,\infty]$, which correspond to the small black hole region and the large black hole region, respectively. The unstable region $v\in[\alpha,\beta]$ is referred to as the spinodal region. In the unstable region, the small and large black hole phase can coexist. Furthermore, we note that $\frac{\partial P(v,T_0)}{\partial v}<0$ in two stable regions but $\frac{\partial P(v,T_0)}{\partial v}>0$ in the unstable region. Mathematically, the two extreme points $\alpha$ and $\beta$ are determined by $\frac{\partial P(v,T_0)}{\partial v}|_{v=\alpha}=\frac{\partial P(v,T_0)}{\partial v}|_{v=\beta}=0$, and the inflection point at $v=v_0$ is determined by $\frac{\partial^2 P(v,T_0)}{\partial v^2}=0$.

Some works have suggested the Melnikov method is well suited for studying chaotic behavior of the black hole within the extended thermodynamic phase space[34-37]. Hence, we shall investigate the chaos phenomenon of the Kerr-AdS black hole under periodic thermal perturbations with the help of the Melnikov method.

**3. Temporal chaos in a spinodal region**

Here, the effect of a weak temporally periodic perturbation shall be considered, when the Kerr-AdS black hole is quenched to the spinodal region. According to the standard procedure, we need to make use of the Kerr-AdS black hole equation of state to construct the Hamiltonian for the fluid flow and derive the Melnikov function containing information about the occurrence of temporal chaos. With respect to the temporal perturbation, we first consider a specific volume $v_0$ (the inflection point) in the spinodal region, which corresponds to an isotherm $T_0$ ($T_0 < T_c$). The weak time-periodic fluctuation of the absolute temperature near $T_0$ can be written as [42]

$$T = T_0 + \varepsilon \gamma \cos(\omega t)\cos(M) \quad (11)$$

with $\varepsilon \ll 1$. According to Ref. [34], one can suppose that the black hole flow takes place along the $x$ axis in a finite tube with a unit cross section, which includes a total of a mass $2\pi/q$ of the black hole in a volume $(2\pi/q)v_0$. Here, $q$ is a positive constant.

From Ref. [34], $x_0$ can be denoted as the Eulerian coordinate of a reference system. Thus, the mass $M$ of a column of the Kerr-AdS black hole with the unit cross section between the reference $x_0$ and a general Eulerian coordinate $x$ can be expressed as

$$M = \int_{x_0}^{x} \rho(\xi,t)d\xi, \quad (12)$$

where $\rho(x,t)$ denotes the black hole density at the position $x$ and the time $t$. It should be pointed out that $\rho(x(M,t),t)^{-1} = x_M(M,t) = v$, where $v$ represents the specific volume.

Now, let us consider the thermodynamic phase transition displayed in the one dimensional thermal compressible Kerr-AdS black hole flow, which can be depicted in Lagrangian coordinates via the following system:

$$\frac{\partial v}{\partial t} = \frac{\partial u}{\partial M}, \quad \frac{\partial u}{\partial t} = \frac{\partial \tau}{\partial M}. \quad (13)$$

Here, $u$ stands for the velocity and $\tau$ corresponds to the stress tensor. According to Korteweig's theory, $\tau$ is defined as

$$\tau = -P(v,T) + \mu \frac{\partial u}{\partial M} - A \frac{\partial^2 v}{\partial M^2} . \tag{14}$$

After substituting (14) into (13), we can obtain

$$\frac{\partial^2 x}{\partial t^2} = -\frac{\partial p(v,T)}{\partial M} + \mu \frac{\partial^3 x}{\partial t \partial M^2} - A \frac{\partial^3 x}{\partial M^4} . \tag{15}$$

Let $\tilde{M} \to qM$, $\tilde{t} \to qt$, $\tilde{x} \to qx$, $\mu \to \varepsilon \mu_0$, Eq. (15) can be recast as

$$\frac{\partial^2 x}{\partial t^2} = \frac{\partial P(v,T)}{\partial M} + \varepsilon \mu_0 q \frac{\partial^3 x}{\partial t \partial M^2} - Aq^2 \frac{\partial^4 x}{\partial M^4} . \tag{16}$$

Here, $\varepsilon$ and $\mu_0$ are positive constants. For convenience, "~" has been omitted in Eq. (16). Then, the mass-volume constraint can be written as

$$\int_0^{2\pi} v(M,t) dM = 2\pi v_0 . \tag{17}$$

It is not hard to construct the Hamiltonian of the system, which has the following form

$$H = \frac{1}{\pi} \int_0^{2\pi} \left[ \frac{u^2}{2} + F(v,T) + \frac{Aq^2}{2} \left( \frac{\partial v}{\partial M} \right)^2 \right] dM , \tag{18}$$

where $A$ is a positive constant and $F(v,T)$ is given by [35]

$$F(v,T) = -\int_{v_0}^{v} \bar{P}(\zeta,T) d\zeta . \tag{19}$$

Notice that $\bar{P}(\zeta,T) = P(\zeta,T) dV/d\zeta$ is an effective equation of state obtained by replacing $\zeta$ in terms of the thermodynamic volume $V = \pi \zeta^3/6$ before performing the integral. Considering that $v = v_0$ is the infection point in the spinodal region, thus we have $P_{vv}(v_0,T_0) = 0$, $P_v(v_0,T_0) > 0$, and $P_{vvv}(v_0,T_0) < 0$. At the equilibrium point $(v_0,T_0)$ with $v = v_0$ and $u = 0$, $v$ and $u$ can be expand in Fourier cosine and sine series on $[0,2\pi]$, respectively. By the use of Eq. (17), we can obtain

$$v(M,t) = x_M(M,t)$$
$$= v_0 + x_1(t) \cos M + x_2(t) \cos 2M + x_3(t) \cos 3M + \ldots,$$

$$u(M,t) = x_t(M,t)$$
$$= u_1(t)\sin M + u_2(t)\sin 2M + u_3(t)\sin 3M + \ldots \quad (20)$$

For the sake of carrying out the perturbation analysis, we need to expand $\bar{P}(v,T)$ near the equilibrium point $(v_0,T_0)$ in a Taylor series and keep terms to third order. Thus, one has[35]

$$\bar{P}(v,T) = \bar{P}(v_0,T_0) + \bar{P}_v(v_0,T_0)(v-v_0) + \bar{P}_T(v_0,T_0)(T-T_0) + \frac{1}{2}\bar{P}_{vv}(v_0,T_0)(v-v_0)^2$$
$$+ \bar{P}_{vT}(v_0,T_0)(v-v_0)(T-T_0) + \frac{\bar{P}_{vvv}(v_0,T_0)(v-v_0)^3}{3!}$$
$$+ \frac{\bar{P}_{vvT}(v_0,T_0)(v-v_0)^2(T-T_0)}{2}.$$

(21)

Considering that $\bar{P}_{TT}(v_0,T_0)$, $\bar{P}_{vTT}(v_0,T_0)$, and $\bar{P}_{TTT}(v_0,T_0)$ are equal to be zero, hence they have been ignored in the above expression. By substituting Eqs. (19)-(21) into Eq. (18), the Hamiltonian can be rewritten as

$$H_2 = \frac{u_1^2}{2} + \frac{u_2^2}{2} - (\frac{\pi}{4}T_0 - \frac{48J^2}{v_0^5})x_1^2 - (\frac{\pi}{4}T_0 - \frac{48J^2}{v_0^5})x_2^2$$
$$- \frac{120J^2}{v_0^6}x_1^2 x_2 + \frac{90J^2}{v_0^7}x_1^4 + \frac{360J^2}{v_0^7}x_1^2 x_2^2 + \frac{90J^2}{v_0^7}x_2^4$$
$$+ \frac{Aq^2 x_1^2}{2} + 2Aq^2 x_2^2 - \frac{\pi v_0}{2}\varepsilon\gamma\cos(\omega t)x_1 - \frac{\pi}{4}\varepsilon\gamma\cos(\omega t)x_1 x_2,$$

(22)

where $(x_1, x_2)$ and $(u_1, u_2)$ stand for the positions and velocities of the first two modes. Consequently, it is not difficult to drive the corresponding equations of motion, which are as follows

$$\dot{x}_1 = u_1,$$

$$\dot{x}_2 = u_2,$$

$$\dot{u}_1 = -\frac{\partial H_2}{\partial x_1} - \varepsilon\mu_0 q u_1$$
$$= (\frac{\pi}{2}T_0 - \frac{96J^2}{v_0^5})x_1 + \frac{240J^2}{v_0^6}x_1 x_2 - \frac{360J^2}{v_0^7}x_1^3 - \frac{720J^2}{v_0^7}x_1 x_2^2 - Aq^2 x_1$$
$$+ \frac{\pi v_0}{2}\varepsilon\gamma\cos(\omega t) + \frac{\pi}{4}\varepsilon\gamma\cos(\omega t)x_2 - \varepsilon\mu_0 q u_1,$$

$$\dot{u}_2 = -\frac{\partial H_2}{\partial x_2} - 4\varepsilon\mu_0 q u_2$$

$$= (\frac{\pi}{2}T_0 - \frac{96J^2}{v_0^5})x_2 + \frac{120J^2}{v_0^6}x_1^2 - \frac{720J^2}{v_0^7}x_1^2 x_2$$

$$- \frac{360J^2}{v_0^7}x_2^3 - 4Aq^2 x_2 + \frac{\pi}{4}\varepsilon\gamma\cos(\omega t)x_1 - 4\varepsilon\mu_0 q u_2.$$

(23)

By setting $z = (x_1, x_2, u_1, u_2)^T$, the above equations can be organized in a compact form, namely,

$$\dot{z}(t) = g_0(z) + \varepsilon g_1(z,t).$$  (24)

Here, the small perturbation $g_1(z,t)$ is periodic in time. The unperturbed system ($\varepsilon = 0$) is given by

$$\dot{z}(t) = g_0(z).$$  (25)

By linearizing the unperturbed system about $z = 0$, one can obtain

$$\dot{z}_A(t) = \mathbf{A}z_A(t),$$  (26)

where the Jacobian matrix $\mathbf{A}$ reads [42]

$$\mathbf{A} = \begin{pmatrix} 0 & 0 & 1 & 0 \\ 0 & 0 & 0 & 1 \\ -Aq^2 + \varphi & 0 & -\varepsilon\mu_0 q & 0 \\ 0 & -4Aq^2 + \varphi & 0 & -4\varepsilon\mu_0 q \end{pmatrix}$$

(27)

with $\varphi = \frac{\pi}{2}T_0 - \frac{96J^2}{v_0^5}$. Its eigenvalues are as follows

$$\lambda_{1,2} = \frac{-\varepsilon\mu_0 q}{2} \pm \frac{1}{2}[\varepsilon^2 \mu_0^2 q^2 - 4(Aq^2 - \frac{\pi T_0}{2} + \frac{96J^2}{v_0^5})]^{\frac{1}{2}},$$

$$\lambda_{3,4} = -2\varepsilon\mu_0 q \pm [4\varepsilon^2 \mu_0^2 q^2 - (4Aq^2 - \frac{\pi T_0}{2} + \frac{96J^2}{v_0^5})]^{\frac{1}{2}}.$$

(28)

If the parameters satisfy the following constraints

$$\frac{\frac{\pi T_0}{2} - \frac{96J^2}{v_0^5}}{4A} < q^2 < \frac{\frac{\pi T_0}{2} - \frac{96J^2}{v_0^5}}{A}$$  (29)

and $\varepsilon$ ($>0$) is small enough, then the first mode is unstable with $\lambda_1 > 0$ and $\lambda_2 < 0$ but the second and higher modes are stable.

Next, let us analyze the unperturbed ($\varepsilon = 0$) system $\dot{z}(t) = g_0(z)$ existing a two-dimensional invariant symplectic manifold, which contains a homoclinic orbit and connecting the origin to itself. The corresponding analytical form can be written as [38]

$$z_0(t) = \begin{pmatrix} \left(\dfrac{\beta^2 v_0^7}{180 J^2}\right)^{\frac{1}{2}} \mathrm{sech}(\beta t) \\ 0 \\ -\beta^2 \left(\dfrac{v_0^7}{180 J^2}\right)^{\frac{1}{2}} \mathrm{sech}(\beta t)\tanh(\beta t) \\ 0 \end{pmatrix}$$

(30)

with

$$\beta = \sqrt{\varphi - Aq^2}\,. \tag{31}$$

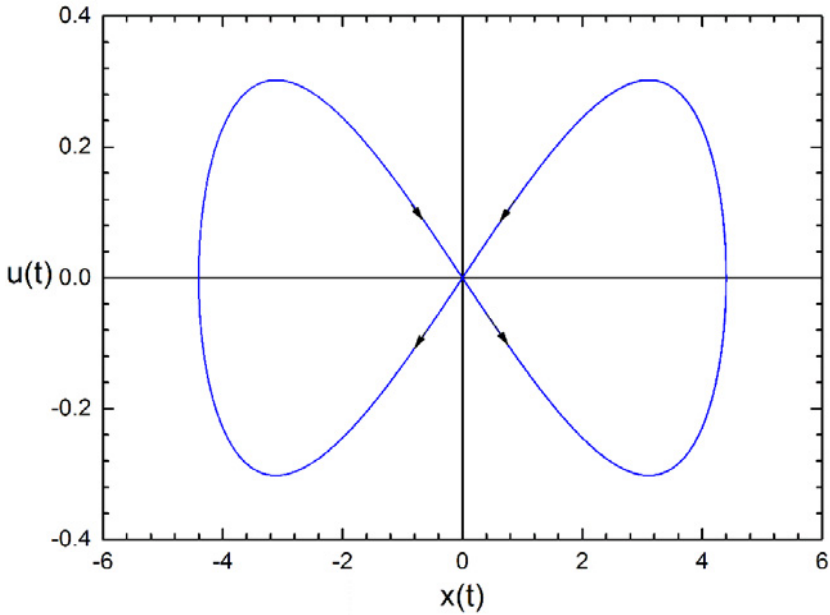

Fig. 2. (color online) The homoclinic orbit of the unperturbed system with $T = 0.02894 < T_c$. The parameters are set to $J = 1$, $q = 1$, and $A = 0.01$. These arrows stand for the time flow direction.

In Fig. 2, we show the phase portrait corresponding to the homoclinic orbit of the unperturbed system, which agrees with the analytical expression (30). From this figure, we can see that the homoclinic orbit possesses the two branches, which are two wings of the butterfly-like orbit, respectively. According to Eq. (30), it is easy to deduce that $z_0$ tends to a saddle point (i.e., the origin point) as $t \to \pm\infty$.

After introducing the time-periodic perturbation (10) (i.e., $\varepsilon \neq 0$), the above homoclinic orbit may be destroyed, which causes that the chaotic behavior of the system is possible. In the light of the Melnikov method, the Melnikov function for the present perturbed system can be computed by using the following formula [41]

$$M(t_0) = \int_{-\infty}^{+\infty} g_0^T(z_0(t-t_0)) \mathbf{J}_{n=2} g_1(z_0(t-t_0),t) dt \tag{32}$$

with

$$\mathbf{J}_{n=2} = \begin{pmatrix} 0 & 1 & 0 & 0 \\ -1 & 0 & 0 & 0 \\ 0 & 0 & 0 & 1 \\ 0 & 0 & 0 & 0 \end{pmatrix}.$$

(33)

Here, $g_0(z_0(t-t_0))$ and $g_1(z_0(t-t_0),t)$ have the following forms

$$g_0(z_0(t-t_0)) = \begin{pmatrix} -\beta^2 \left(\dfrac{v_0^7}{180J^2}\right)^{\frac{1}{2}} \operatorname{sech}[\beta(t-t_0)] \tanh[\beta(t-t_0)] \\ 0 \\ \beta^2 \left(\dfrac{\beta^2 v_0^7}{180J^2}\right)^{\frac{1}{2}} \operatorname{sech}[\beta(t-t_0)] - \dfrac{360J^2}{v_0^7} \left(\dfrac{\beta^2 v_0^7}{180J^2}\right)^{\frac{3}{2}} \operatorname{sech}^3[\beta(t-t_0)] \\ \dfrac{2}{3} v_0 \beta^2 \operatorname{sech}^2[\beta(t-t_0)] \operatorname{sech}^2[\beta(t-t_0)] \end{pmatrix}$$

(34)

and

$$g_1(z_0(t-t_0),t) = \begin{pmatrix} 0 \\ \dfrac{\pi}{2} \varepsilon \gamma v_0 \cos(\omega t) + \varepsilon \mu_0 q \beta^2 \left(\dfrac{v_0^7}{180J^2}\right)^{\frac{1}{2}} \operatorname{sech}[\beta(t-t_0)] \tanh[\beta(t-t_0)] \\ 0 \\ \dfrac{\pi}{4} \left(\dfrac{\beta^2 v_0^7}{180J^2}\right)^{\frac{1}{2}} \operatorname{sech}[\beta(t-t_0)] \varepsilon \gamma \cos(\omega t) \end{pmatrix}.$$

(35)

After straightforward calculations, we find that the Melnikov function can be written

in the following form

$$M(t_0) = N\omega\gamma \sin(\omega t_0) - q\mu_0 I \qquad (36)$$

with

$$N = \left(\frac{v_0^7}{180J^2}\right)^{\frac{1}{2}} \frac{\pi^2 v_0}{2} \text{sech}(\frac{\pi\omega}{2\beta}), \quad I = \frac{v_0^7 \beta^3}{270J^2} \qquad (37)$$

Note that $M(t_0)$ has simple zeros at $N\omega\gamma \sin(\omega t_0) - q\mu_0 I = 0$ only when

$$\left|\frac{q\mu_0 I}{N\omega\gamma}\right| \leq 1. \qquad (38)$$

Hence, it is not hard to conclude that the sufficiently small temporal perturbation may cause the occurrence of the chaos behavior due to the temporal thermal fluctuation. What is more, Eq. (38) can be translated into a critical value for the perturbation parameter $\gamma$, which reads

$$\gamma_c = \frac{2\sqrt{5}q\mu_0 v_0^{\frac{5}{2}} \beta^3 \cosh(\frac{\pi\omega}{2\beta})}{45J\pi^2 \omega}. \qquad (39)$$

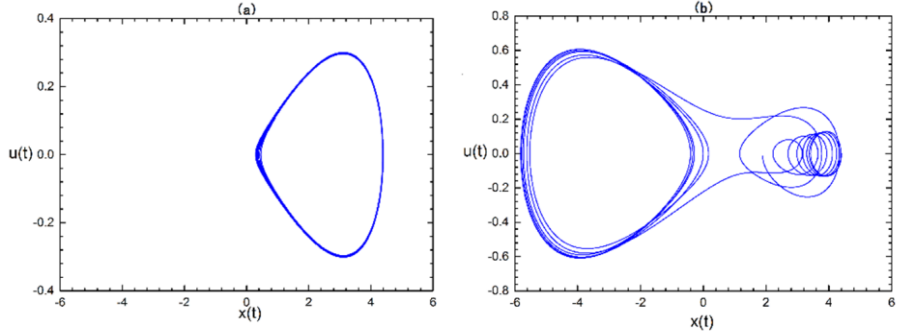

Fig. 3. (color online) Temporal evolution in the phase space of velocity vs displacement for the perturbed system in the specific temperature $T = 0.02894 < T_c$ :(a) $\gamma = 0.0188 < \gamma_c$ :(b) $\gamma = 5 > \gamma_c$. Parameters are set to $J = 1$, $q = 1$, $A = 0.01$, $\varepsilon = 0.001$, $\omega = 0.01$, and $\mu_0 = 0.1$. The initial conditions are chosen as a fixed point $((\sqrt{\beta^2 v_0^7/(180J^2)}, 0)$ in the right branch of the homoclinic orbit.

We note that the small temporal perturbation with $\gamma > \gamma_c$ guarantees the transversal intersection between unstable and stable manifolds, which may give rise to the emergence of the Smale horseshoe chaotic motion[40]. In order to check this analytical condition on the chaotic threshold, numerical results for different values of $\gamma$ are shown in Fig. 3. For convenience, both $x_2$ and $u_2$ have been fixed as zero. Figs. 3(a) displays normal trajectories of the system in the presence of the small temporal perturbation (for $\gamma < \gamma_c$). Figure 3(b) shows the occurrence of chaotic motion for $\gamma > \gamma_c$.

From Eq. (39), one can conclude that the critical value $\gamma_c$ depends on the value of the angular momentum $J$. In Fig. 4, one can clearly see that $\gamma_c$ decreases as the value of the angular momentum $J$ increases. This means that the larger $J$ makes the occurrence of the chaos behavior easier under the time-periodic thermal perturbation.

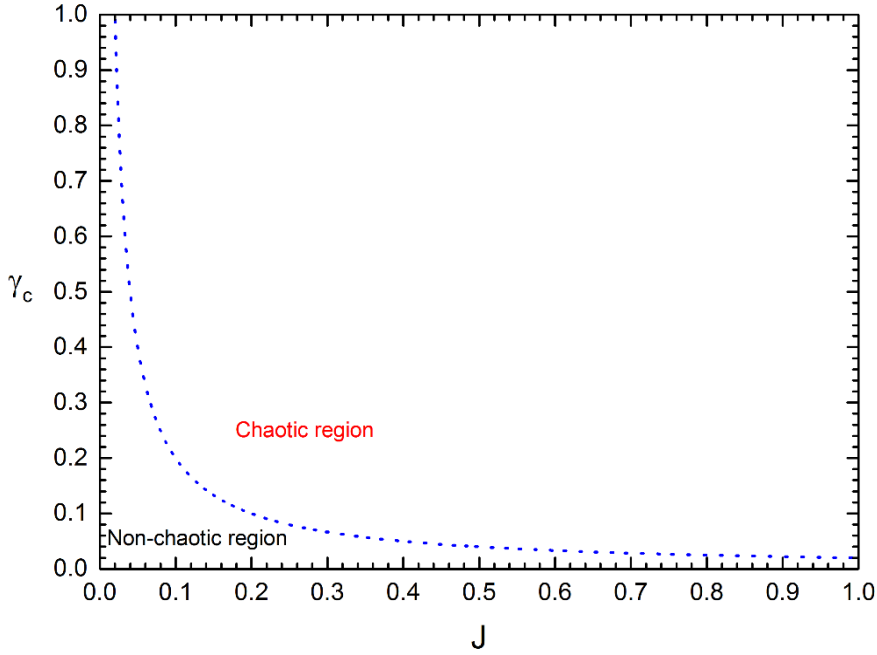

Fig. 4. (color online) Dependence of the critical value $\gamma_c$ on the angular momentum $J$ for the Kerr-AdS black hole. Other parameters are set as in Fig. 2.

## 4. Spatial chaos in the equilibrium state

In this section, let us consider the effect of the small spatially periodic perturbation in the equilibrium configuration with an absolute temperature ($T_0 < T_c$) of the form [42]

$$T = T_0 + \varepsilon \cos(qx). \quad (40)$$

On the basis of the Korteweig theory, the stress tensor can be written as

$$\tau = -p(v,T) - Av'', \tag{41}$$

where $p(v,T)$ satisfies the Kerr-AdS black hole equation of state in Eq. (8), $A > 0$ is a constant, and the symbol $'$ stands for $\frac{d}{dx}$. It should be pointed out that, for a static equilibrium without body forces, one can obtain $\frac{d\tau}{dx} = 0$ so as to $\tau = const = -B$. Furthermore, $B$ is the ambient pressure as $|x| \to \infty$. Based on this fact, Eq. (41) can be changed into the following form

$$v'' = B - p(v,T). \tag{42}$$

Here, the constant $A$ has been set to $A = 1$ as Refs. [34,35].

First, we discuss the unperturbed system ($T = T_0$). For arbitrary temperature $T = T_0 < T_c$, the nonlinear systems in Eq. (42) exist three fixed points, which are $v_1$, $v_2$ and $v_3$, respectively. In the light of the magnitude of the ambient pressure $B$, Eq. (42) can generate three different types of portraits in the $v - v'$ phase plane:

**Case 1.** The ambient pressure $B$ lies in the range $P_0 < B < P(\beta, T_0)$. $P_0$ is the phase transition pressure. Then, the values $v_1$, $v_2$ and $v_3$ so that $P(v_1, T_0) = P(v_2, T_0) = P(v_3, T_0) = B$ are displayed in Fig. 5(a), meanwhile the portrait of Eq. (42) in the $v - v'$ phase plane is displayed in Fig. 5(b).

**Case 2.** The ambient pressure $B$ lies in the range $P(\alpha, T_0) < B < P_0$. Then, the values $v_1$, $v_2$ and $v_3$ so that $P(v_1, T_0) = P(v_2, T_0) = P(v_3, T_0) = B$ are presented in Fig. 6(a), meanwhile the corresponding $v - v'$ phase plane is presented in Fig. 6(b).

**Case 3.** The ambient pressure $B$ is equal to the phase transition pressure $P_0$, i.e., $B = P_0$. In this case, the values $v_1$, $v_2$ and $v_3$ so that $P(v_1, T_0) = P(v_2, T_0) = P(v_3, T_0) = P_0$ are exhibited in Fig. 7(a), meanwhile the corresponding $v - v'$ phase plane is exhibited in Fig. 7(b).

From these figures, one can conclude that the present unperturbed system exists a homoclinic orbit for both case 1 and case 2 while it has a heteroclinic orbit for the final case. It is obvious that the above characteristics allow us to calculate the Melnikov function for these orbits.

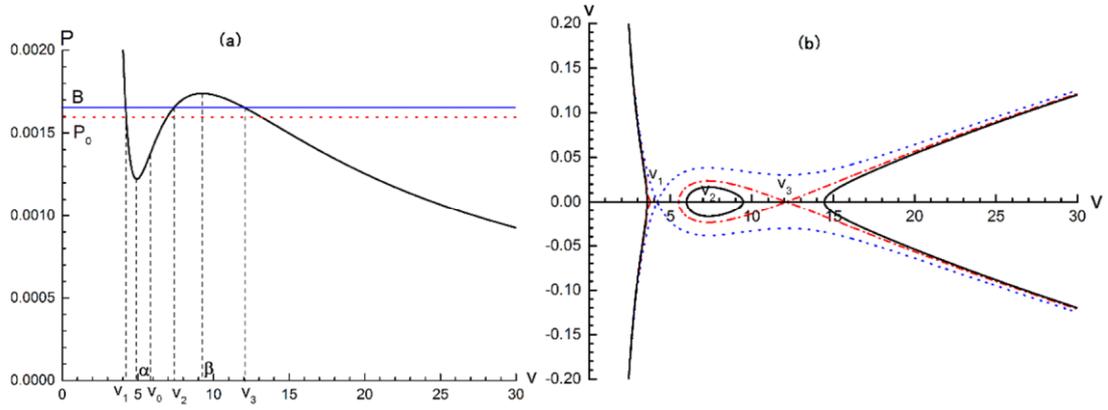

Fig. 5. (color online) Case 1: (a) $B$ and $v_1$, $v_2$, $v_3$ in $P-v$ diagram; (b) $v-v'$ phase portrait.

One can see a homoclinic orbit (read dash dot line) connecting $v_3$ to itself.

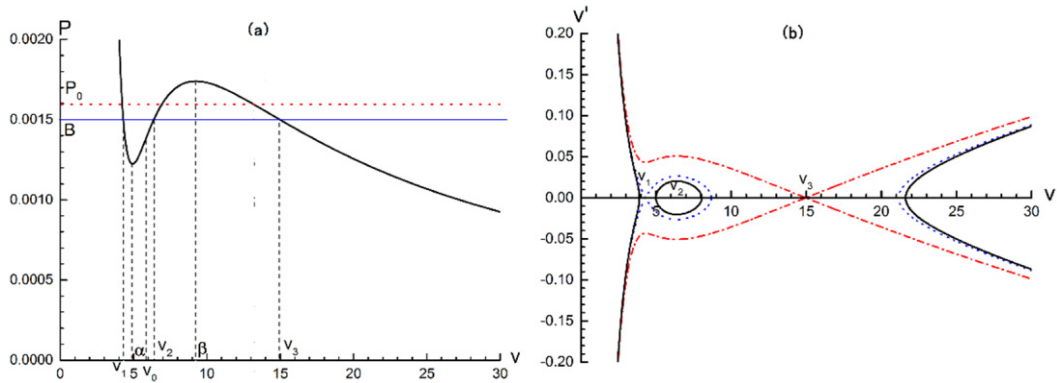

Fig. 6. (color online) Case 2: (a) $B$ and $v_1$, $v_2$, $v_3$ in $P-v$ diagram; (b) $v-v'$ phase portrait.

One can see a homoclinic orbit (blue dash dot line) connecting $v_1$ to itself.

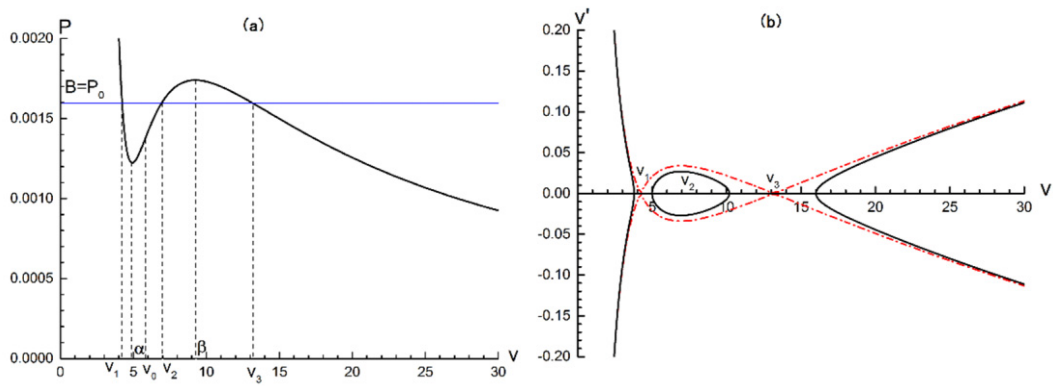

Fig. 7. (color online) Case 3: (a) $B$ and $v_1$, $v_2$, $v_3$ in $P-v$ diagram; (b) $v-v'$ phase portrait.

One can see a heteroclinic orbit (read dash dot line) connecting $v_1$ to $v_3$.

After adding a small spatial perturbation expressed in Eq. (40), we can rewrite

the Eq. (43) for perturbed system as follows:

$$v'' = B - p(v, T_0) - \frac{\varepsilon \cos(qx)}{v}. \tag{43}$$

Then, the Melnikov function can be expressed as [42]

$$M(x_0) = \int_{-\infty}^{+\infty} F^T(Z(x-x_0)) \mathbf{J}_{n=1} G(Z(x-x_0), x) dx \tag{44}$$

with

$$\mathbf{J}_{n=1} = \begin{pmatrix} 0 & 1 \\ -1 & 0 \end{pmatrix}. \tag{45}$$

Let $v' = w$, then Eq. (43) can be changed into two first-order equations, which read

$$v' = u,$$

$$w' = B - P(v, T_0) - \frac{\varepsilon \cos(qx)}{v}.$$

$$\tag{46}$$

As in the previous section, the general solutions of homoclinic or heteroclinic orbit can be written as

$$Z(x-x_0) = \begin{pmatrix} v_0(x-x_0) \\ w_0(x-x_0) \end{pmatrix}. \tag{47}$$

Then, the expressions of the $F$ and $G$ functions are respectively given by

$$F(Z(x-x_0)) = \begin{pmatrix} w_0(x-x_0) \\ B - p(v_0(x-x_0), T_0) \end{pmatrix}, \tag{48}$$

and

$$G(Z(x-x_0)) = \begin{pmatrix} 0 \\ -\frac{\cos(qx)}{v_0(x-x_0)} \end{pmatrix}. \tag{49}$$

By introducing a new variable $X = x - x_0$, the Melnikov function can be reduced to

$$M(x_0) = -L \cos(qx_0) + K \sin(qx_0) \tag{50}$$

with

$$L = \int_{-\infty}^{+\infty} \frac{w_0(X) \cos(qX)}{v_0(X)} dX, \quad K = \int_{-\infty}^{+\infty} \frac{w_0(X) \sin(qX)}{v_0(X)} dX. \tag{51}$$

From Eq. (50), one can find that $M(x_0)$ always exists simple zeros for any given values of $L$ and $K$, which signals that spatial chaos can occur in the thermodynamic system of the Kerr-AdS black hole under the spatially periodic thermal perturbation. This result is in agreement with other AdS black holes [34-37]. In Fig. 8-Fig. 10, the numerical solutions of the perturbed dynamical equation (46) for these three cases are plotted in the $v-v'$ plane, respectively. The corresponding initial configures are chosen to be the homoclinic orbit or heteroclinic orbit. These figures clearly show that there exists indeed the spatial chaos under the spatially periodic thermal perturbation.

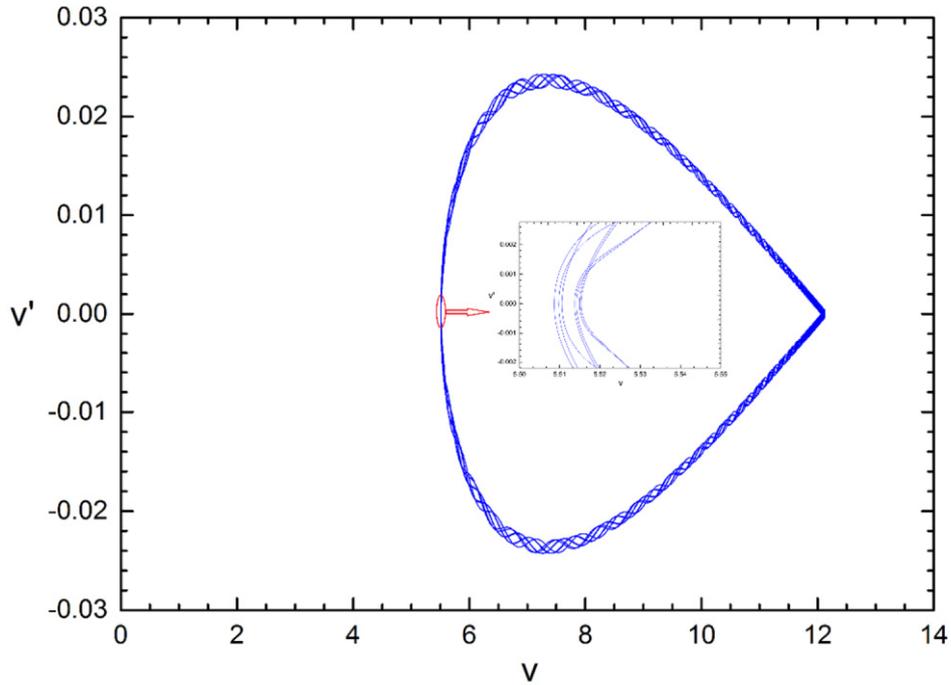

Fig. 8.  (color online) Portrait of the perturbed equation in $v-v'$ phase plane for case 1.

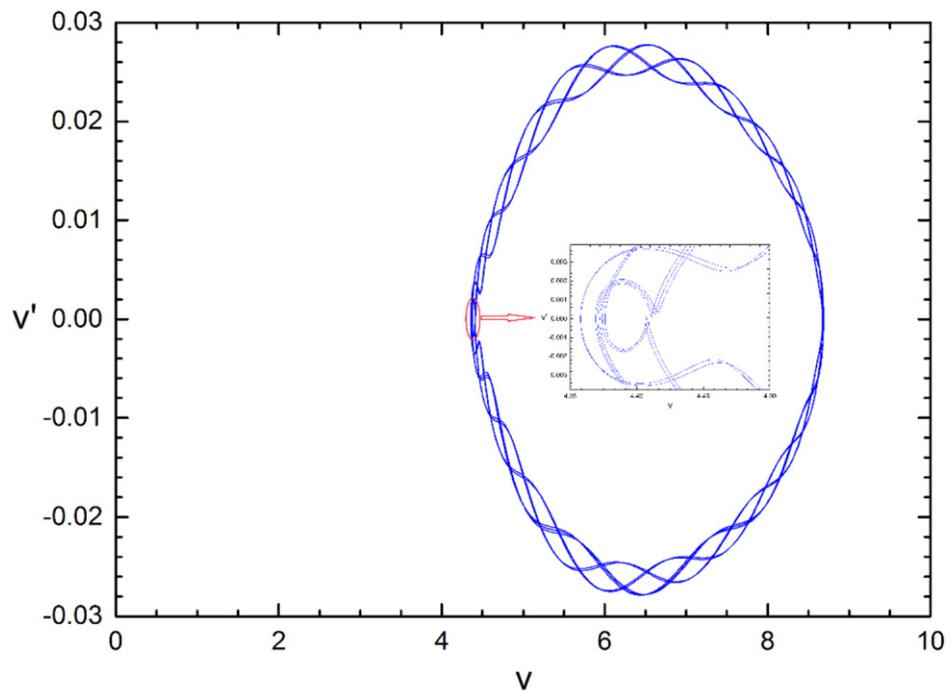

Fig. 9.  (color online) Portrait of the perturbed equation in $v-v'$ phase plane for case 2.

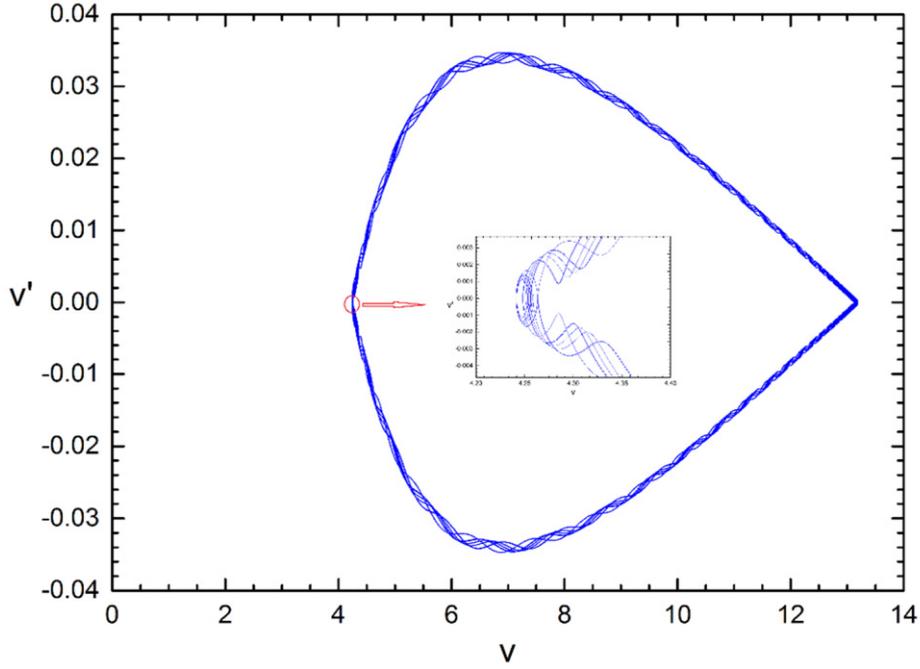

Fig. 10. (color online) Portrait of the perturbed equation in $v-v'$ phase plane for case 3.

## 5. Summary

In summary, we have investigated the occurrence of chaotic behavior under temporally and spatially periodic perturbations in the unstable spinodal region of the Kerr-AdS black hole within the extended phase space. The perturbed Hamiltonian system corresponding to the motion of the Van der Waals system in the spinodal region, following from the Kerr-AdS black hole equation of state, was derived and found to have nonlinear terms giving rise to homoclinic or heteroclinic orbits in the extended phase space. Analysis of the zeroes of the appropriate Melnikov functions provides information about the onset of chaos phenomena in the extended thermodynamic phase space. Under the temporal perturbation in the spinodal region, we displayed that the zeros of the Melnikov function give a critical value of the temporal perturbation $\gamma_c$ for the occurrence of the temporal chaos. Only when the perturbation amplitude $\gamma$ is larger than the critical value $\gamma_c$, the chaotic behavior can emerge. Our results showed that the larger $J$ makes the occurrence of the chaos behavior easier under the time-periodic thermal perturbation. Moreover, we also consider the spatial thermal perturbations that is periodic in space. It is found that the spatial chaos can occur in the small/large black hole equilibrium configuration when the system is suffered from the spatial thermal perturbation. Since the Melnikov function $M(x_0)$ always possesses simple zeros, the occurrence of the spatial chaos is independent of the perturbation amplitude.

In the present work, the Melnikov method has been utilized to probe chaotic characteristics so as to reveal the deep connection between a neutral rotating AdS black hole and the Van der Waals fluid system. To our knowledge, this is the first study on thermal chaos in the rotating AdS black hole. We hope that our results can help to further understand the occurrence of the chaos with respect to the phase picture of such rotating AdS black holes within the extended phase space. For simplicity, we have restricted ourselves to the neutral rotating AdS black hole in the four-dimensional spacetime. Extensions to higher dimension and charged rotating AdS black holes are straightforward. It is also interesting to see if thermal chaos can appear in these cases. Let us leave the exploration of such phenomena for future work.

**Acknowledgments**

The author thanks Professor Songbai Chen of Hunan Normal University for helpful suggestions. This work was supported by the National Natural Science Foundation of China under Grant Nos. 11604121 and 11875126.